\documentclass[english,aps,prl,twocolumn,superscriptaddress,longbibliography]{revtex4-2}

\usepackage{amssymb}
\usepackage{natbib}
\usepackage{epstopdf}
\usepackage{braket}
\usepackage{physics}
\usepackage{mathrsfs}
\usepackage{upgreek}
\usepackage{todonotes}
\usepackage[colorlinks=true, allcolors=blue]{hyperref}
\usepackage{lineno}

\usepackage{xcolor}
\usepackage{graphicx}
\usepackage{color}
\usepackage{comment}
\usepackage[normalem]{ulem}
\usepackage{amsmath}
\usepackage{lipsum}
\usepackage{siunitx}

\definecolor{mycolor}{RGB}{0, 0, 0}

\newcommand{\RomanNumeralCaps}[1]
{\MakeUppercase{\romannumeral #1}}
\begin{document}
\raggedbottom
\title{Spin-Valley Protected Kramers Pair in Bilayer Graphene}

\author{Artem~O.~Denisov}
\email{adenisov@phys.ethz.ch}
\author{Veronika~Reckova}
\author{Solenn~Cances}
\author{Max~J.~Ruckriegel}
\author{Michele~Masseroni}
\author{Christoph~Adam}
\author{Chuyao~Tong}
\author{Jonas~D.~Gerber}
\author{Wei~Wister~Huang}
\affiliation{Laboratory for Solid State Physics, ETH Zurich, CH-8093 Zurich, Switzerland}
\author{Kenji~Watanabe}
\affiliation{Research Center for Functional Materials, National Institute for Materials Science, 1-1 Namiki, Tsukuba 305-0044, Japan}
\author{Takashi~Taniguchi}
\affiliation{International Center for Materials Nanoarchitectonics,
National Institute for Materials Science, 1-1 Namiki, Tsukuba 305-0044, Japan}
\author{Thomas~Ihn}
\author{Klaus~Ensslin}
\affiliation{Laboratory for Solid State Physics, ETH Zurich, CH-8093 Zurich, Switzerland}
\affiliation{Quantum Center, ETH Zürich, CH-8093 Zürich, Switzerland}
\author{Hadrien~Duprez}
\email{hadrien.duprez@polytechnique.edu}
\affiliation{Laboratory for Solid State Physics, ETH Zurich, CH-8093 Zurich, Switzerland}

% Set the distance between line numbers and text
%\setlength\linenumbersep{0.2em} 

% Enable line numbers and place them on the outer edge of each column
%\modulolinenumbers[1] % Number lines in steps of 5 (optional)
%\linenumbers
%\switchlinenumbers

\maketitle
\textbf{
The intrinsic valley degree of freedom makes bilayer graphene (BLG) a unique platform for semiconductor qubits. The single-carrier quantum dot (QD) ground state exhibits a two-fold degeneracy, where the two states that constitute a Kramers pair, have opposite spin and valley quantum numbers. Because of the valley-dependent Berry curvature, an out-of-plane magnetic field breaks the time-reversal symmetry of this ground state and a qubit can be encoded in the spin-valley subspace. The Kramers states are protected against known spin- and valley-mixing mechanisms because mixing requires a simultaneous change of both quantum numbers. Here, we fabricate a tunable QD device in Bernal BLG and measure a spin-valley relaxation time for the Kramers states of ${38~\mathrm{s}}$, which is two orders of magnitude longer than the ${0.4~\mathrm{s}}$ measured for purely spin-blocked states. We also show that the intrinsic Kane-Mele spin-orbit splitting enables a Kramers doublet single-shot readout even at zero magnetic field with a fidelity above ${99\%}$. If these long-lived Kramers states also possess long coherence times and can be effectively manipulated, electrostatically defined QDs in BLG may serve as long-lived semiconductor qubits, extending beyond the spin qubit paradigm.
}

\twocolumngrid

%\section{Introduction}

\begin{figure*}[tbh!]
	\includegraphics[width=1.45\columnwidth]{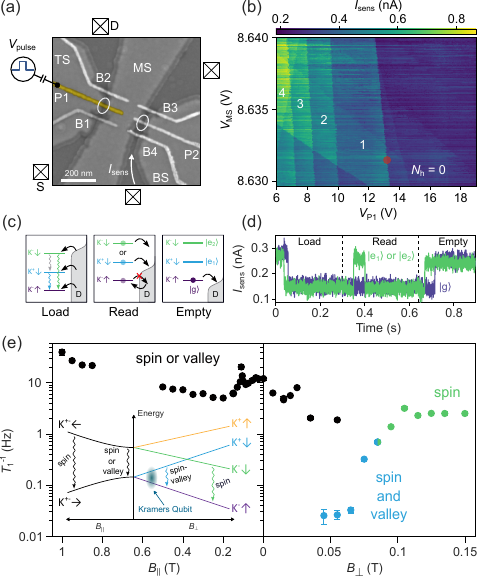}
	\caption{ \textbf{Device and pulsing protocol used to determine relaxation times.}~(a)~SEM false-color image of the device. All labeled metallic gates -- including barrier gates (B1-B4), plunger gates (P1, P2), split gates (TS, MS, BS), and the graphite back gate (not shown) -- as well as the source (S) and drain (D) ohmic contacts, are DC-biased.. The plunger gate P1 is additionally controlled by AC pulses. (b)~Charge stability diagram of the dot probed with a charge sensor current $I_{\mathrm{sens}}$ as a function of the middle-split gate $V_{\mathrm{MS}}$ and plunger gate $V_{\mathrm{P1}}$ voltages. $N_{\mathrm{h}}$ indicates the number of hole carriers inside the QD. The red mark indicates the point on the gate-gate diagram where all measurements were performed. (c)~Three-step pulse readout protocol. Schematic energy diagrams for the three lowest states of the single hole in the BLG QD at $B_{\perp}=\SI{40}{mT}$. (d)~Single-shot readout of the hole state. Typical time-traces of the sensor current $I_{\mathrm{sens}}$ in response to a three-level pulse. For the green trace, the hole is declared to be in one of the excited states (ESs) by the characteristic \textcolor{mycolor}{current step} during the `Read' phase. For the purple trace, the hole is declared to be in the ground state if no\textcolor{mycolor}{step} is observed during the `Read' phase. (e)~Measured inverse relaxation time $T_{1}^{-1}$ as a function of in-plane $B_{||}$ and out-of-plane $B_{\perp}$ magnetic fields. Marks of different colors at the same magnetic field correspond to different relaxation channels controlled by changing the amplitude of `Load' and `Read' pulses. Data points are presented as mean values $\pm$ the standard deviation of the calculated $T_{1}$ (Methods). Some error bars are smaller than the symbol size of the data point. The inset shows the energy spectrum of a single carrier in the BLG QD plotted as a function of in-plane and out-of-plane magnetic fields.}
	\label{fig_1}
\end{figure*}

\textcolor{mycolor}{Atomically thin semiconductors with a hexagonal crystal lattice serve as a unique host for solid-state qubits~\cite{RevModPhys.95.025003, Chatterjee2021} due to their intrinsic valley degree of freedom. Among them, bilayer graphene stands out as a platform that benefits from carbon's low nuclear spin density and weak spin-orbit coupling, along with the ability to fabricate highly tunable few-carrier quantum dots~\cite{Garreis2024, PRXQuantum.3.020343, Banszerus2023}.} The two-fold valley degeneracy~\cite{McCann_2013, Knothe_2022} provides an opportunity to define a qubit in the two-dimensional subspace spanned by the valley and the spin degrees of freedom~\cite{PhysRevX.4.011034}. One benefit of encoding a qubit using two quantum numbers instead of one is its robustness \textcolor{mycolor}{(protection)} against different types of noise \textcolor{mycolor}{that would otherwise cause it to decay}. While the spin is coupled to surrounding magnetic moments and phonons via spin-orbit coupling (SOC)~\cite{PhysRevApplied.11.044063} or hyperfine interaction~\cite{Camenzind2018}, a valley flip requires a scattering process with a giant momentum swing on the order of the length of the reciprocal lattice vector, which is only known to be caused by short-range perturbations on the length scale of atomic defects. \textcolor{mycolor}{Given the low disorder reached in encapsulated graphene~\cite{doi:10.1126/science.1244358}}, the valley relaxation times were demonstrated to be at least an order of magnitude longer than spin relaxation times~\cite{Garreis2024} within a single device. One can speculate that the simultaneous flip of both spin and valley~\cite{doi:10.1021/acs.nanolett.3c01779}, thus, is a rare second-order process.

%BLG is not the only material with a valley degree of freedom~\cite{RevModPhys.85.961}, but, in contrast to silicon, the energy of valley states in BLG is tunable by a perpendicular magnetic field $B_{\perp}$~\cite{doi:10.1021/acs.nanolett.0c04343}. Additionally, valleys in silicon are oriented normal to the plane of the carrier gas, making them susceptible to spatial fluctuations in the interface potential~\cite{RevModPhys.85.961}, whereas the valleys in graphene are oriented in the plane, potentially leading to much less adverse effects of the BLG/hBN interface.

A \textcolor{mycolor}{remarkable} property of BLG is a large out-of-plane valley $g$-factor ($g_{\mathrm{v}}$), which originates from the finite Berry curvature around the $K$-points~\cite{PhysRevB.98.155435} and can be tuned using the electric field across a wide range of values from 10 to 100~\cite{doi:10.1021/acs.nanolett.0c04343}. At zero magnetic field, the four-fold single-particle degeneracy is already lifted by Kane--Mele spin--orbit interaction~\cite{PhysRevLett.95.226801}. %This interaction can be imagined as a constant out-of-plane magnetic field seen by a charge carrier, with its sign changing depending on the product of the valley and spin quantum numbers.
It separates the energy of the two Kramers doublets ($|K^{-}{\uparrow}\rangle$, $|K^{+}{\downarrow}\rangle$) and ($|K^{-}{\downarrow}\rangle$, $|K^{+}{\uparrow}\rangle$) by a spin-orbit gap, which has been previously measured indirectly~\cite{Kurzmann2021, Banszerus2021, Banszerus2022} and directly~\cite{Banszerus2021, Duprez2024} to be $\Delta_{\mathrm{SO}}\approx \qtyrange[range-units = single , range-phrase = -]{40}{80}{\micro eV}\mathrel{\widehat{=}} \qtyrange[range-units = single , range-phrase = -]{10}{20}{GHz}$. \textcolor{mycolor}{Comparable spin-orbit splittings are observed between Kramers pairs in semiconducting transition metal dichalcogenide (TMD) multilayer systems, such as MoS$_2$~\cite{doi:10.1021/acs.nanolett.3c01779}.} By increasing $B_{\perp}$, time-reversal symmetry is broken resulting in an energy splitting of both Kramers pairs. Previous studies showed coherent oscillations of a spin--valley qubit formed in a carbon nanotube (CNT)~\cite{Laird2013} but yielded short relaxation times $T_{1}<\SI{10}{\micro s}$~\cite{Laird2013, PhysRevLett.102.166802} and decoherence times $T_2^{\mathrm{(echo)}}=\SI{65}{ns}$ presumably attributed to high levels of static disorder and charge noise~\cite{Laird2013, PhysRevLett.102.166802, Kuemmeth2008}. In contrast, encapsulated BLG is known to be atomically flat, and with significantly slower valley mixing rate $\Delta_{\mathrm{{K^{+}K^{-}}}}<\SI{2}{neV}$~\cite{Garreis2024}, promising \textcolor{mycolor}{longer relaxation times}.

In this work, we perform relaxation time measurements of a single-hole quantum dot formed in BLG with small ($\sim\mathrm{mT}$) or zero applied magnetic field. We find the relaxation time $T_{1}$ for simultaneous spin–valley relaxation to be as long as $T^{\mathrm{(sv)}}_{1}\approx30~\mathrm{s}$ which is two orders of magnitude longer than for pure spin relaxation $T^{\mathrm{(s)}}_{1}\approx0.4~\mathrm{s}$. Additionally, we demonstrate high fidelity readout at zero magnetic field where relaxation happens between two Kramer’s doublets and requires either a spin or a valley flip.

We form a tunable QD device, electrostatically defined in Bernal BLG, using two layers of overlapping gates depicted in Fig.~\ref{fig_1}a (see Methods for fabrication details). The lower gate layer serves the purpose of confining the BLG charge carriers to two conducting channels via opening the band gap in BLG underneath the top (TS), middle (MS), and bottom (BS) split gates, and placing the Fermi energy into the gap in these regions~\cite{https://doi.org/10.1002/aelm.202200510}. The upper gate layer of narrow finger gates is used to precisely define the QDs within the channels. We form the signal and the sensor QDs underneath their respective plunger gates P1 and P2. The dot-to-reservoir couplings can be tuned individually by pairs of barrier gates (B1, B2) and (B3, B4), respectively. We tune the dot into the single-hole regime as shown in Fig.~\ref{fig_1}b, where the sensor current $I_{\mathrm{sens}}$ is shown as a function of middle split and plunger gate voltages $V_{\mathrm{MS}}$ and $V_{\mathrm{P1}}$. From the width of the $0-1$ charge transition, we determine the electron temperature to be $T_{\mathrm{e}} \approx 30~\mathrm{mK}$.

 %We attribute the set of parallel discrete steps to the boundaries of the regions of the charge stability diagram with a constant hole occupation $N_{\mathrm{h}}=0,~1,~2,\ldots$ as indicated in the figure. We identify the first hole transition around $V_{\mathrm{P1}}\approx\SI{13}{V}$ (marked by the red dot), as no charge transition line appears at higher plunger gate voltages. Additional lines with drastically different couplings to MS and P1 gates correspond to unintended dots formed due to nearby charge inhomogeneities.

\begin{figure*}[tbh!]
	\includegraphics[width=1.85\columnwidth]{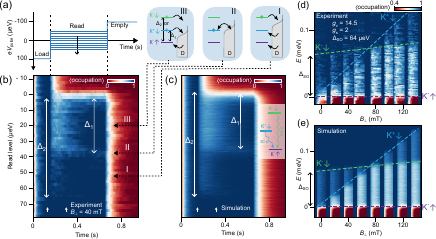}
	\caption{\textbf{Single-shot energy spectroscopy.}~(a)~Calibration of the `Read' level and the excited state spectroscopy. The `Read' level is varied while the `Load' and `Empty' are kept constant. (b)~The digitized sensor current (representing dot occupation) averaged over $\sim140$ single-shot traces as a function of the `Read' level shift relative to the GS at $B_{\perp}=40~\mathrm{mT}$. Insets sketch the relative position of the dot's energy levels and the $E_\mathrm{F}$ of the lead for three different regimes: (\RomanNumeralCaps{3}) readout from both ESs, (\RomanNumeralCaps{2}) resonant jumps between the lower energy ES and the lead, (\RomanNumeralCaps{1}) readout from the higher energy ES. $\Delta_{1}$ and $\Delta_{2}$ denote the energy splittings between the GS and the two ESs respectively. The sensor signal is digitized to represent the dot occupation, such that current offsets due to the sensor drift are compensated. (c)~Monte-Carlo simulation of the single-shot readout averaged over 10,000 shots for each `Read' level. We use the simplest model with three states, as shown on the sketch, where both of the ESs can only relax to the GS with $T_{1}^{\mathrm{(s)}}=0.5~\mathrm{s}$ and $T_{1}^{\mathrm{(sv)}}=30~\mathrm{s}$ --- two values measured independently. (d)~Excited state spectroscopy. The beginning of the `Read' \textcolor{mycolor}{phase (spanning the time window marked with white arrows in (b)) is shown for $B_{\perp}=[0,~20,~40,~55,~80,~100,~115,~135]~\mathrm{mT}$.} The `Read' level shift relative to the GS. The dashed lines show the evolution of the two lower energy ESs in a single-hole BLG QD: $|K^{-}{\downarrow}\rangle$ (green) and $|K^{+}{\downarrow}\rangle$ (blue), with $\Delta_{\mathrm{SO}}=64~\mathrm{\mu eV}$ and $g_{\mathrm{v}}=14.5$. (e)~Similar data as in (d) but for Monte-Carlo simulated single-shot readout of the QD. The averaged \textcolor{mycolor}{step} maps do not reveal solid evidence for the existence of the fourth state $|K^{+}{\uparrow}\rangle$ presumably due to the short \textcolor{mycolor}{relaxation time} of the hole in this state. However, we could observe all four non-degenerate states and their predicted magnetic field dependence using more precise tunneling rate spectroscopy measurements~\cite{Duprez2024} as shown in the Supplementary Figure~1. \textcolor{mycolor}{Scale bar is shared with (d).}}
	\label{fig_2}
\end{figure*}

\section{Three-level single-shot readout}
We perform the standard Elzerman three-level single-shot readout~\cite{Elzerman2004, Morello2010, PRXQuantum.3.020343} to extract the relaxation times (see Methods for the details). We apply voltage pulses to the plunger gate P1, while monitoring the real-time charge occupancy of the dot using the charge sensor. The concept of the readout is sketched in Fig.~\ref{fig_1}c: if a hole remains in the excited state after the 'Load' and during the `Read' phase without relaxing, then, at some random time governed by the tunneling-out rate, it will tunnel to the leads and thereafter the ground state of the dot will be occupied again through a tunneling-in process. This in-and-out tunneling manifests itself as a \textcolor{mycolor}{step} in the sensor current shown in Fig.~\ref{fig_1}d (green trace). 
%In contrast, no `blip' is observed when the hole was initially loaded into the ground state, or when it relaxed before it could tunnel out during the `Read' phase (purple trace). 
By determining the ratio of the number of shots with a single \textcolor{mycolor}{step} to the total number of successfully loaded shots and discarding error traces (see Methods for postselection procedure), we extract the relaxation time by fitting the exponential decay as a function of the load time~\cite{PhysRevLett.106.156804}. 
%We use a two-threshold algorithm to determine the dot occupation from the sensor signal and discard error traces such as those with multiple `blips' or those where no hole is loaded during the `Load' phase. We assume that these errors are uncorrelated (see Methods).

Figure~\ref{fig_1}e shows the main results of this paper: the measured relaxation times $T_{1}$ are plotted as a function of small out-of-plane and in-plane magnetic fields. As we will prove later by excited state spectroscopy measurements, at $B_{\perp}<\SI{80}{mT}$, the transition between the two lowest energy states in the single-particle spectrum are spin--valley blocked, so the resulting $T_1$ times can be interpreted as spin--valley relaxation times $T_{1}^{\mathrm{(sv)}}$ (blue points in the figure). We observed long relaxation times $T_{1}^{\mathrm{(sv)}}\approx \SI{30}{s}$.
This is in tune with the intuition that, a simultaneous flip of both quantum numbers is a very rare second-order relaxation process. Note that the range in magnetic field where we can measure this spin--valley relaxation time is bounded from below by $B_{\perp}\gtrsim\SI{40}{mT}$ and from above by $B_{\perp}\lesssim\SI{70}{mT}$. We notice, that at lower energy splittings, the fidelity of Elzerman readout drastically drops, affected by finite temperature and charge noise~\cite{Keith_2019}. We would like to note that similarly long $T_{1}$ times were measured in GaAs~\cite{Camenzind2018} in the limit close to $B=0$. In both cases the times are probably rather limited by the required measurement time and sample stability than by the two-level system itself.

%At zero magnetic field in Fig.~\ref{fig_1}e, each of the Kramers pairs is two-fold degenerate and the pairs are split from each other by the SO gap $\Delta_{\mathrm{SO}}=\SI{64}{\micro eV}$.
At low perpendicular magnetic fields, by choosing the `Read' level such that $E_\mathrm{F}$ in the lead is inside the gap between the Kramers pairs, we can perform high-fidelity readout of the pair state and measure relaxation times between two doublets which are blocked by either spin or valley flip (black points in Fig.~\ref{fig_1}e). As a result, the resulting relaxation time is dominated by the fastest of two channels and can be expressed as $T_{1}~=~(1/T_{1}^{\mathrm{(s)}}~+~1/T_{1}^{\mathrm{(v)}})^{-1}$. 
%It is, however, not possible to disentangle the two contributions to the total relaxation rate since doublets of states are degenerate at $B_{\perp}=0$. 

Applying a perpendicular magnetic field lifts the Kramers pairs' degeneracies, and enables us to distinguish the spin and spin--valley relaxation rates by accurately placing the $E_\mathrm{F}$ of the lead between the corresponding states. Around $B_{\perp}\approx80~\mathrm{mT}$ in Fig.~\ref{fig_1}e, two excited states cross, namely $|K^{-}{\downarrow}\rangle$ and $|K^{+}{\downarrow}\rangle$. For $B_{\perp}>80~\mathrm{mT}$, the transition between the two lowest states, $|K^{-}{\downarrow}\rangle$ and $|K^{-}{\uparrow}\rangle$, is purely spin-blocked, meaning that the measured $T_1$ times can be identified with spin relaxation times $T_{1}^{\mathrm{(s)}}$ (green data points). We measured $T_{1}^{\mathrm{(s)}}\approx\SI{0.4}{s}$ which is two orders of magnitude shorter than the spin--valley relaxation time, but still one order of magnitude longer than spin relaxation times previously measured at higher magnetic fields $B_{\perp}>\SI{1.5}{T}$~\cite{Garreis2024, PRXQuantum.3.020343}. This finding is consistent with phonon spin relaxation mediated by SOC and/or hyperfine coupling~\cite{PhysRevLett.100.046803, Camenzind2018}. Such long spin $T_1$ times are also consistent with the fact the QD is highly decoupled from the leads with the intrinsic tunneling rate as low as $\Gamma_{\mathrm{out}}\approx \SI{15}{Hz}$ (see Methods for $\Gamma_{\mathrm{in/out}}$ extraction). Additionally, we find no relaxation hotspot~\cite{Yang2013, PhysRevApplied.11.044063} where two different valley states cross (at $B_{\perp}=\SI{80}{mT}$) in Fig.~\ref{fig_1}e. This is consistent with the low inter-valley mixing term measured in BLG~\cite{Garreis2024, Duprez2024}. In Methods section, we explicitly rule out the argument that the observed long spin-valley relaxation times could be interpreted as fully dominated by pure spin relaxation.

In Fig.~\ref{fig_1}e left we also show the non-monotonic dependence of the $T_1$-time as a function of in-plane magnetic field $B_\parallel$ that does not break the Kramers degeneracy but slowly increases the energy splitting of the Kramers pairs. Starting from $T_{1}\approx\SI{100}{ms}$ at zero magnetic field, the relaxation time doubles around $B_\parallel=\SI{200}{mT}$ and then decreases monotonically at higher fields, reaching $T_{1}\approx\SI{25}{ms}$ at $B_\parallel=\SI{1}{T}$. We also observe two insignificant relaxation hotspots around $B_\parallel=\SI{110}{mT}$ and $B_{\perp}=\SI{20}{mT}$. These hotspots leave no traces in spectroscopy measurements (see Supplementary Figure 2).

\section{Single-shot excited state spectroscopy}
To support the identification of the different relaxation channels discussed above, we carefully probe the state spectrum of our QD by standard `Read' window calibration~\cite{Morello2010, Yang2013}. In Fig.~\ref{fig_2}b, we graph the occupation probability of the QD estimated from $\sim 140$ single shot traces, varying the `Read' level position relative to the alignment between the ground state and the Fermi energy in the leads. The `Load' ($eV_{\mathrm{load}}=~\SI{95}{\micro eV}$) and `Empty' ($eV_{\mathrm{empty}}=~\SI{-95}{\micro eV}$) levels are kept constant during this process (see the sketch in Fig.~\ref{fig_2}a). %The sensor signal is digitized to represent the dot occupation, such that current offsets due to the sensor drift are compensated.
%When the `Read' level is negative (GS above $E_\mathrm{F}$ in the leads), the dot becomes empty right after entering the `Read' phase regardless of the hole state. In this regime, the average dot occupation during the `Read' phase is constant and equal to zero at times longer than $1/\Gamma_{\mathrm{out}}$. In contrast, at very high positive `Read' levels above $\sim70~\mathrm{\mu eV}$, the dot stays filled during the entire `Read' phase, since all ESs that the hole could potentially be loaded into remain below $E_\mathrm{F}$ in the leads. For energies in-between, each single-shot trace shows either a constant dot occupation of 1, or a single `blip' where the occupation flips from 1 to 0 and back, indicating that the hole was in an ES at the beginning of the `Read' phase and tunneled out before relaxing. 
The finite probability of having a \textcolor{mycolor}{step} in the sensor trace reduces the average dot occupation at the beginning of the `Read' phase as seen from Fig.~\ref{fig_2}b (light blue region). We can clearly distinguish three different regimes here: (\RomanNumeralCaps{1}) Low density of \textcolor{mycolor}{steps} between $\SI{35}{\micro eV}$ and $\SI{65}{\micro eV}$ as we read out only the higher energy ES $|K^{-}{\downarrow}\rangle$, (\RomanNumeralCaps{3}) high density of \textcolor{mycolor}{steps} between $0$ and $\SI{35}{\micro eV}$ as we readout both of the ESs and (\RomanNumeralCaps{2}) where the density of \textcolor{mycolor}{steps} decays very slowly over time, indicating resonant back-and-forth jumps of the hole between the lower energy ES $|K^{+}{\downarrow}\rangle$ and the lead as their energies align. 
%This is also a hallmark of the relaxation time between the spin--valley blocked lower energy ES and the GS, being much longer than the tunneling time $1/\Gamma_{\mathrm{out}}+1/\Gamma_{\mathrm{in}}\approx\SI{115}{ms}$. The energy scales $\Delta_1$ and $\Delta_2$ indicated in Fig.~\ref{fig_2}b denote the separation of the two ESs from the GS.
In Fig.~\ref{fig_2}d we plot the beginning of the `Read' phase (within a time-interval indicated by white arrows in Fig.~\ref{fig_2}b) at different out-of-plane magnetic fields. We clearly observe the predicted~\cite{Knothe_2022} crossing between two ESs around $B_{\perp}\approx80~\mathrm{mT}$. 
At zero magnetic field, the GS $|K^{-}{\uparrow}\rangle$ and the first ES $|K^{+}{\downarrow}\rangle$ form a Kramers pair and $B_{\perp}$ splits it with $\Delta_{1}=(g_{\mathrm{v}}+g_{\mathrm{s}})\mu_{\mathrm{B}}B_{\perp}$. Relaxation of the second ES $|K^{-}{\downarrow}\rangle$ into the GS is spin-blocked and evolves in $B_{\perp}$ as $\Delta_{2}=g_{\mathrm{s}}\mu_{\mathrm{B}}B_{\perp}+\Delta_{\mathrm{SO}}$. The extracted $g_{\mathrm{v}}=14.5$ ( we checked that $g_{\mathrm{s}}=2$ independently, see Supplementary Figure 1), and the Kane--Mele SO gap $\Delta_{\mathrm{SO}}=64\pm4~\si{\micro eV}$ agree with the previously measured values~\cite{Duprez2024, Kurzmann2021, Banszerus2021, Banszerus2022}.

We can qualitatively compare the data with the results of a simple three state Monte-Carlo model of the single-shot readout~\cite{Morello2010} (see Supplementary Information, section \RomanNumeralCaps{5} for details) with $T_{1}^{\mathrm{(s)}}=0.5~\mathrm{s}$ and $T_{1}^{\mathrm{(sv)}}=30~\mathrm{s}$. Figure~\ref{fig_2}c shows the results of the simulations which aim to qualitatively match the experimental \textcolor{mycolor}{step} map in Fig.~\ref{fig_2}b. 
%We use the simplest model with three states, where both of the ESs can only relax to the GS with $T_{1}^{\mathrm{(s)}}=0.5~\mathrm{s}$ and $T_{1}^{\mathrm{(sv)}}=30~\mathrm{s}$ --- two values measured independently. The tunneling rates and energy splittings used for the simulation are extracted from Fig.~\ref{fig_2}b. 
%Note, that such a single map does not provide quantitative insight about the relaxation rates. However, from the fact that `blips' are visible, we can assert that $T_{1}\gtrsim 1/\Gamma_{\mathrm{out}}+T_{\mathrm{load}}=230~\mathrm{ms}$, where we pick the loading time to satisfy $T_{\mathrm{load}}\gg1/(3\Gamma_{\mathrm{in}})=13~\mathrm{ms}$ to load a hole during each cycle with certainty. 
As expected, the simulation of the $B_{\perp}$-dependence in Fig.~\ref{fig_2}e qualitatively agrees with the experiment in Fig.~\ref{fig_2}d. 

%The averaged `blip' maps (Fig.~\ref{fig_2}b and Fig.~\ref{fig_2}d) do not reveal solid evidence for the existence of the fourth state $|K^{+}{\uparrow}\rangle$ presumably due to the short lifetime of the hole in this state. As a result, our discussion on relaxation times will focus only on the lowest three states. However, we could observe all four non-degenerate states and their predicted magnetic field dependence using more precise tunneling rate spectroscopy measurements~\cite{2311.12949} as shown in the Supplementary Material. 

\begin{figure*}[tbh!]
	\includegraphics[width=1.4\columnwidth]{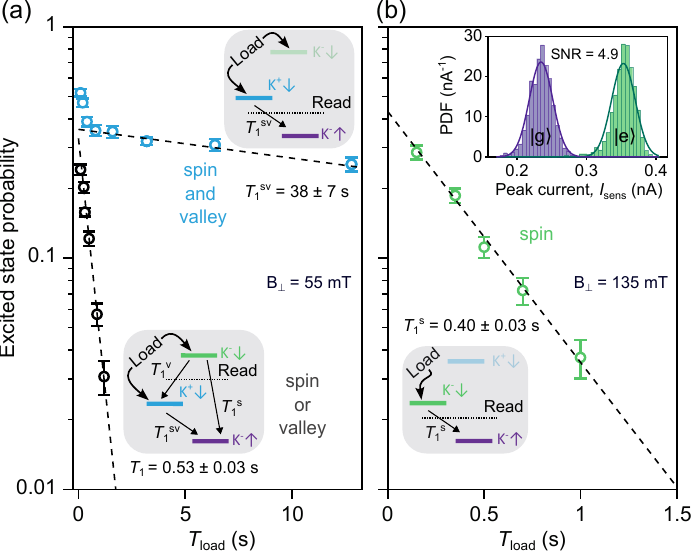}
	\caption{\textbf{Distinguishing different relaxation channels.}~(a)~Exponential decays of the excited state probability at $B_{\perp}=55~\mathrm{mT}$. The upper sketch shows the measurement of the relaxation time from both ESs in parallel (blue circles). Here, the `Read' level is placed below both of the the ESs and above the GS. The lower sketch depicts the relaxation measurements only from the higher energy ES (black circles). Here, the `Read' level is placed between the first and the second ESs. Dashed lines show the exponential fits with outlined $T_{1}$ times. (b)~Exponential decay of the ES probability at $B_{\perp}=135~\mathrm{mT}$. The sketch depicts the standard spin readout protocol: the `Load' level is placed between the two ESs, while the `Read' level is placed between the GS and the lower energy ES. The transition between these two states are spin-blocked. Data points in (a) and (b) are presented as mean values $\pm$ the confidence interval of a single standard deviation (Methods). Inset shows a probability density function (PDF) histograms of the maximum values of the sensor current in the `Read' phase extracted separately from the shots where the state at the beginning of the `Read' phase is identified as the ES (green) and the GS (purple). The solid lines are Gaussian fits obtained separately for each histogram.}
	\label{fig_4}
\end{figure*}

\section{Identifying different relaxation channels}

We start the discussion on how we access all three types of blockade in Fig.~\ref{fig_1}e with perpendicular magnetic fields beyond the crossing $B_{\perp}>80~\mathrm{mT}$. Here the relaxation between the first ES $|K^{-}{\downarrow}\rangle$ and the GS $|K^{-}{\uparrow}\rangle$ is spin-blocked as depicted in Fig.~\ref{fig_4}b. %A hole is equally likely to be loaded into either one of these states by positioning $E_\mathrm{F}$ between the first and second ES during the `Load' phase. 
During the `Read' phase, the $E_\mathrm{F}$ of the lead (represented by the dotted line in Fig.~\ref{fig_4}a and b) is placed between the GS and first ES to facilitate spin-to-charge conversion. The exponential decay of the spin-down probability in Fig.~\ref{fig_4}b yields long $T^{\mathrm{(s)}}_{1}=0.40\pm0.03~\mathrm{s}$~\cite{Garreis2024, PRXQuantum.3.020343}.
 
At lower magnetic fields $B_{\perp}<80~\mathrm{mT}$, the ESs swap their order, and the relaxation between the now lower energy ES $|K^{+}{\downarrow}\rangle$ and the GS $|K^{-}{\uparrow}\rangle$ becomes spin-and-valley blocked. %Unfortunately, the decreasing energy splittings pose an experimental challenge in precisely controlling both the `Load' and `Read' levels simultaneously due to the slow charge drift of the device. 
Due to the presence of occasional charge jumps, we find it easier to load a hole into both ESs rather than accurately placing the `Load' level between the ESs. The lower sketch in Fig.~\ref{fig_4}a illustrates the readout of the higher energy ES $|K^{-}{\downarrow}\rangle$ which requires a spin or valley flip to relax to the GS or the lower energy ES, respectively. The resulting relaxation rate obtained from the black data points is, therefore, the average of two parallel relaxation channels $T_{1}~=~(1/T^{\mathrm{(s)}}_{1}+1/T^{\mathrm{(v)}}_{1})^{-1}~=~0.53~\pm~0.03~\mathrm{s}$, with a value close to the spin relaxation times and likely dominated by it~\cite{Garreis2024}. 
%This aligns with the previous observation of valley $T_{1}$-times being significantly longer than spin $T_{1}$~\cite{Garreis2024}. 
We want to highlight a key feature in the relaxation measurement of the higher energy ES $|K^{-}{\downarrow}\rangle$: when loading times exceed $T_{\mathrm{load}}\approx2~\mathrm{s}$, all carriers loaded to the higher energy ES relax, leaving it unoccupied. 
 
Alternatively, we can measure relaxation from both the ESs by moving the `Read' level below the lower energy ES, as shown in the upper sketch in Fig.~\ref{fig_4}a. Here, two distinct regimes in the exponential decay of the ES probability (blue data points) are evident: at short loading times, the relaxation rate is fast and attributed to the relaxation from the highest ES, while around $T_{\mathrm{load}}>1~\mathrm{s}$, the second regime of \textcolor{mycolor}{rather} long relaxation becomes apparent. As we have illustrated that the higher energy ES is unoccupied at these loading times, we attribute the long $T_{1}^{\mathrm{(sv)}}=38\pm7~\mathrm{s}$ to the simultaneous spin-valley flip between the lower ES and the GS. The two-orders-of-magnitude difference in relaxation rates for different read levels is striking. To validate our observations and rule out potential thermally activated and flicker noise origin of \textcolor{mycolor}{steps} in Methods.

\section{Conclusions}

The \textcolor{mycolor}{notably} long $T_{1}$ times, together with the demonstrated high-fidelity readout, present exciting opportunities for studying spin and valley physics at low and zero magnetic fields, where it holds particular significance as the energy splittings align with the capabilities of standard microwave equipment operating around $5-12~\mathrm{GHz}$. \textcolor{mycolor}{Performing such measurements in a device based on purified Carbon-12 graphene would allow for a platform free of hyperfine-induced decoherence. Additionally, the significantly weaker spin-mixing component of a spin-orbit coupling in BLG ($\sim5~\mathrm{neV}$)~\cite{PhysRevB.85.115423, Garreis2024}, compared to other established semiconducting platforms ($60~\mathrm{neV}$ for GaAs~\cite{doi:10.1126/science.1183628}, $ 200~\mathrm{neV}$ for Ge~\cite{PhysRevLett.128.126803} and $ 110~\mathrm{neV}$ for silicon metal-oxide-semiconductor~\cite{PhysRevX.9.021028}), suggests that it may not limit spin coherence times in BLG due to charge noise, one of the key factor constraining the performance in conventional platforms~\cite{RevModPhys.95.025003}. Furthermore, replacing the lossy amorphous aluminum oxide between gate layers with dangling-bond-free hBN could potentially reduce charge noise in BLG QD devices~\cite{Wang2022}.}

An open question remains regarding how to coherently manipulate the valley degree of freedom, since the valley magnetic moment is only coupled to the out-of-plane magnetic field. \textcolor{mycolor}{This problem can be addressed by utilizing spin-valley exchange interaction between the (2,0) and (1,1) orbital configurations in a double dot geometry. The resulting Kramers singlet-triplet qubit can be operated similarly to its spin analog~\cite{doi:10.1126/science.1116955, Jirovec2021} leveraging the gate-tunability of the valley $g$-factor, as shown in the Supplementary Figure 4. Additionally, the readout of the Kramers singlet-triplet qubit can be achieved via Pauli spin-valley blockade~\cite{PhysRevLett.128.067702}.} Another approach to manipulating Kramers states relies on a presence of a finite valley mixing term $\propto\tau_{\mathrm{x}}\Delta_{\mathrm{K^{+}K^{-}}}$ in the original QD Hamiltonian due to the breaking of translational symmetry~\cite{PhysRevB.81.195418}. As a result, one can drive spin-valley rotation by exploiting conventional electron spin resonance protocols \textcolor{mycolor}{(see Supplementary Figure 3)} using antennas, micromagnets~\cite{RevModPhys.95.025003, Chatterjee2021} or utilizing the highly anisotropic effective $g$-factor in a curved BLG quantum dot~\cite{PhysRevB.81.195418}, similar to carbon nanotubes~\cite{Laird2013, Pei2012}. \textcolor{mycolor}{Independent of the magnitude of valley mixing}, two-qubit gates between spin-valley qubits can be facilitated through exchange interactions, providing a robust roadmap for quantum information processing.

%In contrast, a widely tunable valley $g$-factor using an electric field is a simple and natural %next step for exploring coherent spin-valley physics in the time domain.

\section{Acknowledgements}
The authors thank Lin Wang, Guido Burkard and Joshua Folk for useful discussions. All authors acknowledge financial support by the European Graphene Flagship Core3 Project, H2020 European Research Council (ERC) Synergy Grant under Grant Agreement 951541, the European Innovation Council under grant agreement
number 101046231/FantastiCOF, NCCR QSIT (Swiss National Science Foundation, grant number 51NF40-
185902). K.W. and T.T. acknowledge support from the JSPS KAKENHI (Grant Numbers 21H05233 and
23H02052) and World Premier International Research Center Initiative (WPI), MEXT, Japan.

\section{Author Contributions Statement}
\subsection{Corresponding authors}
Correspondence and requests for materials should be addressed to A.D., H.D.

\subsection{Author contributions}
T.T. and K.W. grew the hBN. H.D. fabricated the device with inputs from M.M. and M.R..
A.D. acquired and analyzed the data with inputs from S.C, H.D, T.I. and K.E.. 
V.R. performed numerical simulations of the device.
A.D. wrote the manuscript with inputs from all authors.

%\section{Acknowledgements}
\section{Competing interests}
The authors declare no competing financial interests.

\bibliography{bibliography_Kramers}

\section{Methods}

\subsection{Device fabrication}

All the presented data were measured on a single device fabricated on a van der Waals heterostructure stacked by standard dry transfer techniques. The stack consists of a $\SI{35}{nm}$ thick top hBN layer, the Bernal BLG sheet, a $\SI{28}{nm}$ thick bottom hBN, and a graphite back gate layer. The ohmic contacts to the BLG are 1D edge contacts. To form QDs, we utilize the two overlapping layers of Cr/Au (\SI{3}{nm}/\SI{20}{nm}) metallic gates shown in Fig.~\ref{fig_1}a. The upper gate layer consists of finger gates, \SI{20}{nm}-wide and \SI{60}{nm} apart. These gates are deposited on top of a $\SI{26}{nm}$ thick insulating aluminum oxide layer. The widths of the channels defined by the split gates are \SI{40}{nm} for the left channel and \SI{75}{nm} for the right channel, respectively. We tune the dot into the few-hole regime as shown in Fig.~\ref{fig_1}b. We attribute the set of parallel discrete steps to the boundaries of the regions of the charge stability diagram with a constant hole occupation $N_{\mathrm{h}}=0,~1,~2,\ldots$ as indicated in the figure. We identify the first hole transition around $V_{\mathrm{P1}}\approx\SI{13}{V}$ (marked by the red dot), as no charge transition line appears at higher plunger gate voltages. Additional lines with drastically different couplings to MS and P1 gates correspond to unintended dots formed due to nearby charge inhomogeneities.

By applying a large negative voltage $V_{\mathrm{BG}}=\SI{-7.4}{V}$ to the global back gate we induce a large displacement field ($D=\SI{-0.9}{V/nm}$) which opens a band gap in BLG on the order of $\SI{100}{meV}$. This field allows us to significantly decouple the dot from the reservoirs, achieving tunneling rates to the leads of just tens of $\mathrm{Hz}$.

\subsection{Measurement setup}
The sample is mounted on the mixing chamber of a Bluefors LD400 dilution refrigerator, which has a base temperature of $9~\mathrm{mK}$ and an independently extracted electron temperature of around $30~\mathrm{mK}$. All the measurement and control electronics are located at room temperature and are connected to the device via 24 DC lines. Each line is low-pass filtered using RC filters mounted on the mixing chamber plate, with a time constant of approximately $10~\mathrm{\mu s}$. For DC biasing of gates and ohmic contacts, we use in-house built low-noise voltage sources with a cutoff frequency of $7~\mathrm{Hz}$, except for the pulsing gate P1 line, which is left unfiltered and has a cutoff frequency of $1200~\mathrm{Hz}$. The DC plunger gate voltage $V_{\mathrm{P1}}$ is combined with the pulsing voltage $V_{\mathrm{pulse}}$ using a $2~\mathrm{M\Omega}$/$2.7~\mathrm{k\Omega}$ divider at room temperature. The sensor current is amplified using a built in-house  current to voltage converter with a $10~\mathrm{M\Omega}$ feedback resistor, followed by a $\times 100$ analog voltage amplifier and an analog low-pass filter with a cutoff frequency of around $10~\mathrm{kHz}$. Sensor time traces are recorded using an NI-6251 data acquisition card with a sampling frequency of $20~\mathrm{kHz}$.
\subsection{Elzerman sequence}

We start the three-level single-shot readout sequence with the empty dot subjected to the `Load' phase (see Fig.~\ref{fig_1}c): the pulse level shifts all three states below the Fermi energy $E_\mathrm{F}$ of the leads. A hole from the lead can tunnel into any of the three states with almost equal probability~\cite{Duprez2024} of $1/3$. During the `Load' phase of duration $T_{\mathrm{load}}$, the sensor current drops abruptly, indicating that the dot has been filled, as shown in the typical time traces in Fig.~\ref{fig_1}d. In the following `Read' phase (see Fig.~\ref{fig_1}c), the second pulse level puts the excited state of interest above $E_\mathrm{F}$ in the lead while the ground state stays below. If a hole remains in the excited state after loading and during the `Read' phase without relaxing, then, at some random time governed by the tunneling-out rate, it will tunnel to the leads and thereafter the ground state of the dot will be occupied again through a tunneling-in process. This in-and-out tunneling manifests itself as a \textcolor{mycolor}{step} in the sensor current shown in Fig.~\ref{fig_1}d (green trace). In contrast, no \textcolor{mycolor}{step} is observed when the hole was initially loaded into the ground state, or when it relaxed before it could tunnel out during the `Read' phase (blue trace). Once a hole arrives in the ground state, it blocks any further transitions since the number of resonant unoccupied lead states is exponentially suppressed by the low electron temperature $T_{\mathrm{e}}\approx\SI{30}{mK}$~\cite{Duprez2024}. The final pulse level empties the dot by moving all states above $E_\mathrm{F}$ of the leads (see Fig.~\ref{fig_1}c), and after that, the sequence starts again.

\subsection{Postselection of sensor traces}

We digitize current sensor traces using the two-threshold procedure described in~\cite{Duprez2024}. We divide the entire set of $N$ digitized current traces into four groups, $N=(N_{\mathrm{e}}, N_{\mathrm{g}}, N_{\mathrm{nl}}, N_{\mathrm{er}})$. Here, $N_{\mathrm{nl}}$ is the number of traces where a hole was not loaded during the `Load' phase. $N_{\mathrm{g}}$ is the number of traces with zero \textcolor{mycolor}{steps} in the `Read' phase, which we label as a hole being in the ground state. $N_{\mathrm{e}}$ is the number of traces with a single \textcolor{mycolor}{step} during the `Read' phase, which we define as a hole being in the excited state. $N_{\mathrm{er}}$ is for the rest of the traces, which we label as errors, as they exhibit multiple \textcolor{mycolor}{steps} in the `Read' phase. Extended Data Fig.~1 shows typical sensor current traces for each group along with their digitized version. \textcolor{mycolor}{The typical proportions are as follows: $80.89\%$ for the ground state (GS), $9.41\%$ for the excited state (ES), $0.39\%$ for combined errors, and $0.58\%$ for traces that are not loaded.} The two most common types of errors causing multiple \textcolor{mycolor}{step} traces are digitization errors and random thermal/charge noise \textcolor{mycolor}{steps}. The first is due to the fact that the sensor trace doesn't exhibit enough points in both charge states to reliably determine the threshold for digitization. Thermal \textcolor{mycolor}{steps} are exponentially suppressed by low electron temperature, while major charge jumps occur on a very long timescale of minutes and hours. We simply discard all error traces from the dataset, assuming that the errors are uncorrelated with whether the excited or ground state is occupied in the beginning of the `Read' phase. Note that by discarding clearly excited state counts, however marked as thermal \textcolor{mycolor}{step} errors in Extended Data Fig.~1, we underestimate the excited state probability.

\subsection{Calculation of excited state probability}

We perform relaxation time measurements in a regime with slow tunneling rates $\Gamma_{\mathrm{out}} \approx \Gamma_{\mathrm{in}} \sim \SI{15}{Hz}$, which are comparable to the measured $T_1$ at certain magnetic fields. In this regime, the occupation of the excited state after the short loading times $T_{\mathrm{load}} \sim \Gamma_{\mathrm{in}}^{-1}$ is limited not by the relaxation but by the tunneling-in rate. Nevertheless, even for one of the shortest extracted $T_1 \approx \SI{40}{\m s}$, we demonstrate that the decay of the calculated renormalized excited state probability $P^{\mathrm{e}}$ can be fit by a simple exponent. We consider the state readout of two double-degenerate Kramers pairs at $B_{||} = \SI{900}{\m T}$, as shown in Extended Data Fig.~2a. As established in previous measurements~\cite{PhysRevLett.106.156804, PhysRevApplied.11.044063}, the total occupations of the excited, ground, and non-loaded states after the loading time $T_{\mathrm{load}}$ are

\begin{widetext}
\begin{equation}
\begin{split}
     & n^{\mathrm{e}}(T_{\mathrm{load}})=\frac{N_{\mathrm{e}}}{N_{\mathrm{e}}+N_{\mathrm{g}}+N_{\mathrm{nl}}}= \frac{\Gamma_{\mathrm{in}}^{\mathrm{e}}}{\Gamma_{\mathrm{in}}^{\mathrm{e}}+\Gamma_{\mathrm{in}}^{\mathrm{g}}-T_{1}^{-1}} e^{\displaystyle-T_{\mathrm{load}}T_{1}^{-1}}\Big( 1-e^{\displaystyle-T_{\mathrm{load}}(\Gamma_{\mathrm{in}}^{\mathrm{e}}+\Gamma_{\mathrm{in}}^{\mathrm{g}}-T_{1}^{-1})}\Big) \\
  & n^{\mathrm{g}}(T_{\mathrm{load}})=\frac{N_{\mathrm{g}}}{N_{\mathrm{e}}+N_{\mathrm{g}}+N_{\mathrm{nl}}}=\frac{ (\Gamma_{\mathrm{in}}^{\mathrm{g}}-T_{1}^{-1})\Big( 1-e^{\displaystyle-T_{\mathrm{load}}(\Gamma_{\mathrm{in}}^{\mathrm{e}}+\Gamma_{\mathrm{in}}^{\mathrm{g}})}\Big) +\Gamma_{\mathrm{in}}^{\mathrm{e}}\Big( 1-e^{\displaystyle-T_{\mathrm{load}}T_{1}^{-1}}\Big)}{\Gamma_{\mathrm{in}}^{\mathrm{e}}+\Gamma_{\mathrm{in}}^{\mathrm{g}}-T_{1}^{-1}}\\
  & n^{\mathrm{nl}}(T_{\mathrm{load}})=1-n^{\mathrm{e}}-n^{\mathrm{g}}
    \end{split}
\end{equation}

\end{widetext}
where $\Gamma_{\mathrm{in}}^{\mathrm{e}}$ and $\Gamma_{\mathrm{in}}^{\mathrm{g}}$ are tunneling-in rates to the excited and the ground state respectively. We extract the sum of tunneling-in rates $\Gamma_{\mathrm{in}}^{\mathrm{e}}+\Gamma_{\mathrm{in}}^{\mathrm{g}}=4\Gamma_{\mathrm{in}}=\SI{57.6}{Hz}$ and tunneling-out rate $\Gamma_{\mathrm{out}}^{\mathrm{e}}=\Gamma_{\mathrm{out}}^{\mathrm{g}}=\Gamma_{\mathrm{out}}=\SI{12}{Hz}$ by fitting the exponential decay and rise of the average dot occupation during the `Load' and `Empty' phases respectively, as shown in Extended Data Fig.~2a. During the `Read' phase, we measure the excited states with efficiency~\cite{PhysRevLett.106.156804, PhysRevApplied.11.044063}:

\begin{equation}
n_{\mathrm{RO}}=\frac{\Gamma_{\mathrm{out}}^{\mathrm{e}}}{T_{1}^{-1}+\Gamma_{\mathrm{out}}^{\mathrm{e}}}\Big( 1-e^{\displaystyle-T_{\mathrm{read}}(T_{1}^{-1}+\Gamma_{\mathrm{out}}^{\mathrm{e}})}\Big)
\end{equation}

where $T_{\mathrm{read}}$ is the time spent in the `Read' phase and $\Gamma_{\mathrm{out}}^{\mathrm{e}}$ being the intrinsic tunneling rate.

The resulting full and renormalized probabilities are simply products of both efficiencies.

\begin{equation}
\begin{split}
& P^{\mathrm{e}}_{\mathrm{full}}(T_{\mathrm{load}})=\frac{N_{\mathrm{e}}}{N_{\mathrm{e}}+N_{\mathrm{g}}+N_{\mathrm{nl}}}\cdot n_{\mathrm{RO}}=n^{\mathrm{e}} \cdot n_{\mathrm{RO}} \\
& P^{\mathrm{e}}_{\mathrm{renorm}}(T_{\mathrm{load}})=\frac{N_{\mathrm{e}}}{N_{\mathrm{e}}+N_{\mathrm{g}}}\cdot n_{\mathrm{RO}}=\frac{n^{\mathrm{e}}}{n^{\mathrm{e}}+n^{\mathrm{g}}} \cdot n_{\mathrm{RO}} 
\label{eq_3}
\end{split}
\end{equation}

In Extended Data Fig.~2b, we plot the measured renormalized probability $P^{\mathrm{e}}={N_{\mathrm{e}}}/({N_{\mathrm{e}}+N_{\mathrm{g}}})$ along with theoretical curves from \eqref{eq_3}, using $\Gamma_{\mathrm{in}}^{\mathrm{e}}=\SI{22.2}{Hz}$ and $T_{1}=\SI{41}{\m s}$. As expected, at short $T_{\mathrm{load}}$, the full and renormalized occupations drastically differ. However, as long as $T_{\mathrm{load}}>\SI{60}{\m s}\approx3/(\Gamma_{\mathrm{in}}^{\mathrm{e}}+\Gamma_{\mathrm{in}}^{\mathrm{g}})$, the two curves match well, and their downward trend can be successfully approximated by the simple exponential function $P^{\mathrm{e}}_{\mathrm{exp}}(T_{\mathrm{load}})~\sim e^{-T_{\mathrm{load}}/T_{1}}$ as shown with the dashed line. With increasing relaxation time, the renormalized probability becomes closer to the exponent. The best least square weighted exponential fit of the experimental points yields $T_{1}=45\pm 3~\mathrm{ms}$, which is within 10 percent of the value given by the correct renormalized probability expression.

\subsection{Dark counts}

In Extended Data Fig.~3, we rule out potential thermally activated and flicker noise origin (dark counts) of \textcolor{mycolor}{steps} by comparing the \textcolor{mycolor}{step} distribution at significantly different loading times $T_{\mathrm{load}}=0.2~\mathrm{s}$ and $12.8~\mathrm{s}$. The average of all the single-shot $I_{\mathrm{sens}}$ traces identified as the GS (purple) shows a flat behavior during the `Read' phase. 
In contrast, the average of ESs (green) exhibits a bump that rises on a timescale of $1/\Gamma_{\mathrm{out}}$ and decays within $1/\Gamma_{\mathrm{in}}$~\cite{Morello2010}. The exponential decrease in \textcolor{mycolor}{step} density to zero, governed by the tunneling-out rate, allows us to eliminate the charge noise origin of the \textcolor{mycolor}{steps}, as one would anticipate a constant distribution of random charge jumps over the `Read' phase. Another potential source of false-positive counts could be thermally activated \textcolor{mycolor}{steps}. Since we only count single \textcolor{mycolor}{steps}, the distribution might appear similar. However, with average tunneling-in and -out times of $1/\Gamma_{\mathrm{in}/\mathrm{out}}=75~\mathrm{ms}$ and a `Read' duration of $T_{\mathrm{read}}=350~\mathrm{ms}$, we expect a Poissonian distribution $P(k)=\lambda^{k}e^{-\lambda}/k!$ for the probability of a certain number of tunneling events $k$, with an average of $\lambda=4.7$. The probability of having a single \textcolor{mycolor}{step} (corresponding to two tunneling events) is $P(k=2)\approx10\%$, while the probability of having no \textcolor{mycolor}{step} is $P(k=0)\approx1\%$, meaning that the remaining $88\%$ of shots should be discarded as errors. In fact, for $T_{\mathrm{load}}=0.2~\mathrm{s}$, we identify $45.9\%$ of shots with a single \textcolor{mycolor}{step}, $51.9\%$ of shots with no \textcolor{mycolor}{steps}, $0.9\%$ as not-loaded shots, and only $1.3\%$ as multiple-\textcolor{mycolor}{step} errors, effectively ruling out thermal activation as the origin of the \textcolor{mycolor}{steps}.

\subsection{Readout performance}

For quantum information applications, it is crucial to identify the factors limiting the readout performance. We analyze 2000 single-shot traces to extract the histogram of the peak value of the sensor current during the `Read' phase, as illustrated in the inset of Fig.~\ref{fig_4}b. The well-separated Gaussian peaks representing the detection of the GS and the ES, with a signal-to-noise ratio (SNR) of approximately 4.9, result in an electrical readout fidelity exceeding $99.9\%$~\cite{Morello2010, Keith_2019}. The negligible electrical readout error suggests that the readout performance is limited by the spin/valley-to-charge conversion~\cite{Keith_2019}. We find that, despite the deliberate reduction of the tunneling rates, the observed \textcolor{mycolor}{notably} long spin-valley relaxation time still just meets all the minimum requirements~\cite{Keith_2019} for achieving a fault-tolerant $99\%$ readout visibility threshold~\cite{PhysRevApplied.18.064028}: 1) a large energy splitting more than 13-times larger than electron temperature. In our experiment $\Delta_1\approx \SI{55}{\micro eV}\gtrsim 13k_{\mathrm{B}}T\approx \SI{52}{\micro eV}$. 2) a 100-times faster tunnelling-out time than relaxation time. In our experiment, $T_{1}^{\mathrm{(sv)}}=30~\mathrm{s}\gtrsim100/\Gamma_{\mathrm{out}}=8~\mathrm{s}$, and 3) a 12 times larger sampling rate for data acquisition than the reloading rate. In our experiment, $\Gamma_{\mathrm{s}}=10~\mathrm{kHz}>12\Gamma_{\mathrm{out}}=150~\mathrm{Hz}$.

\subsection{Ruling out pure spin relaxation}

Here we explicitly ruling out the argument that the observed long spin-valley relaxation times could be interpreted as fully dominated by pure spin relaxation. Indeed, few seconds to almost-a-minute~\cite{Camenzind2018} spin $T_1$ times have also been observed both in silicon and GaAs quantum dots~\cite{Stano2022}. Moreover, previous studies indeed report strong power law dependencies~\cite{Stano2022}, which could, at first sight, explain the drastic drop in relaxation times as we transition from the pure spin to the spin-valley blockade regime while simultaneously shrinking the energy splitting.

To eliminate this argument, in Extended Data Fig.~4 we plot the data from Fig.~\ref{fig_1}e as a function of the energy splitting between the first excited and the ground states. Additionally, we include the spin relaxation rate $T_{1}^{-1}$ data at higher magnetic fields ($1.5-3~\mathrm{T}$) from G\"achter {\it{et al}}.~\cite{PRXQuantum.3.020343}, measured in an analogous weakly coupled ($\Gamma\sim350~\mathrm{Hz}$) single QD BLG device using the same Elzerman technique. The energy splitting dependence of $T_1$ extracted from both experiments is best described by a power law, $T^{-1}_{1}\propto\Delta E^{2.5}$ (highlighted by the green solid line). The observed power is in line with $\propto\Delta E^{3-7}$, as seen in GaAs and silicon spin qubits, and originates from a combination of electron-phonon relaxation and various spin mixing mechanisms such as hyperfine interaction, spin-orbit coupling, and spin-valley mixing~\cite{Stano2022}. In the case of BLG, neither spin-mixing mechanisms are well-known nor has any theoretical prediction for spin $T_{1}$ dependence on the magnetic field been made while considering the 2D nature of phonons. However, calculations for single-layer graphene show that the power law $T^{-1}_{1}\propto\Delta E^{2-4}$ should not differ much from the mentioned 3D platforms. Taking this into account, if we assume that long spin-valley relaxation is solely dominated by spin relaxation, the trend correlating spin and Kramers points reveals a remarkable power of $\propto\Delta E^{20}$ as marked by the dashed blue line, although fitting a power law on such a restricted energy range should be approached with caution. Hence, a more plausible explanation for such a drastic change in $T_1$ would be the dual protection afforded by simultaneous spin-valley blocking when operating within the Kramers doublet. The continuous connection between the spin-valley (blue dots) and spin (green dots) relaxation data points can be attributed to the finite valley mixing term. Indeed, even small values of $\Delta_{\mathrm{K^{+}K^{-}}}<\SI{2}{\nano eV}\sim\SI{0.5}{MHz}$~\cite{Garreis2024} can provide effective mixing, considering that our shortest loading times are in the tens of milliseconds. Additionally, the spin-or-valley relaxation channel data points align with the pure spin trend, indicating that valley relaxation is significantly longer than spin relaxation, in agreement with previous observations~\cite{Garreis2024}.

\section{Data availability}

The data that support the findings of this study will be made available online through the ETH Research Collection~\cite{ETH_collection} at http://hdl.handle.net/20.500.11850/711208. This includes raw data, analysis scripts, and plotting scripts for figures from the main text, Extended Data, and Supplementary Information.

\clearpage
\onecolumngrid
\begin{center}
    \Large{\textbf{Extended data}}
\end{center}
\setcounter{figure}{0}
\renewcommand{\thefigure}{\arabic{figure}}
\renewcommand{\theequation}{S\arabic{equation}}
\setcounter{secnumdepth}{2}

\begin{figure*}[!h]
\centering
	\includegraphics[width=0.5\columnwidth]{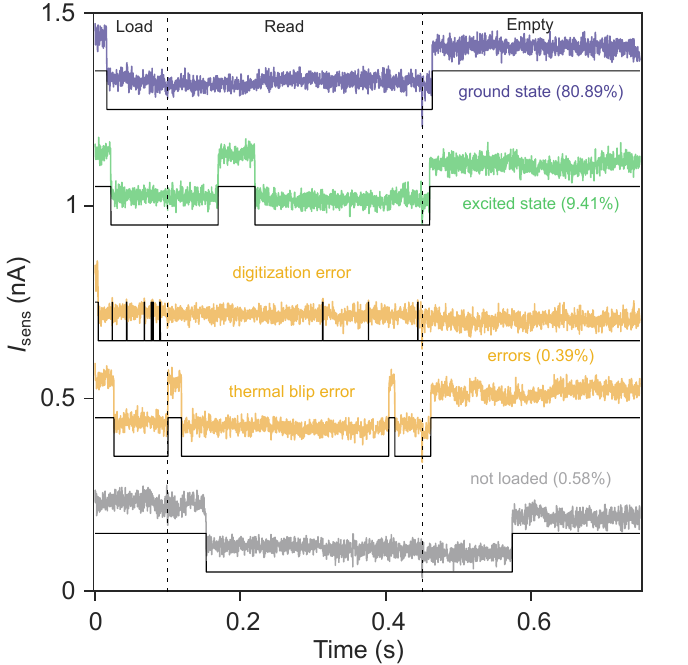}
	\caption{\textbf{Postselection of sensor traces.}~Typical time traces of the sensor current during the three-level pulse. Traces are shifted vertically with an equal offset of $\SI{0.3}{\nano A}$. Black lines show the digitized versions of the traces. By analyzing the digitized signals, we group all traces into four groups: excited states (green), ground states (purple), errors (yellow), or not loaded states (gray). The percentage breakdown for each group is specified. }
	\label{sup_fig_1}
\end{figure*}

\clearpage
\newpage

\begin{figure}[h!]
	\includegraphics[width=0.5\columnwidth]{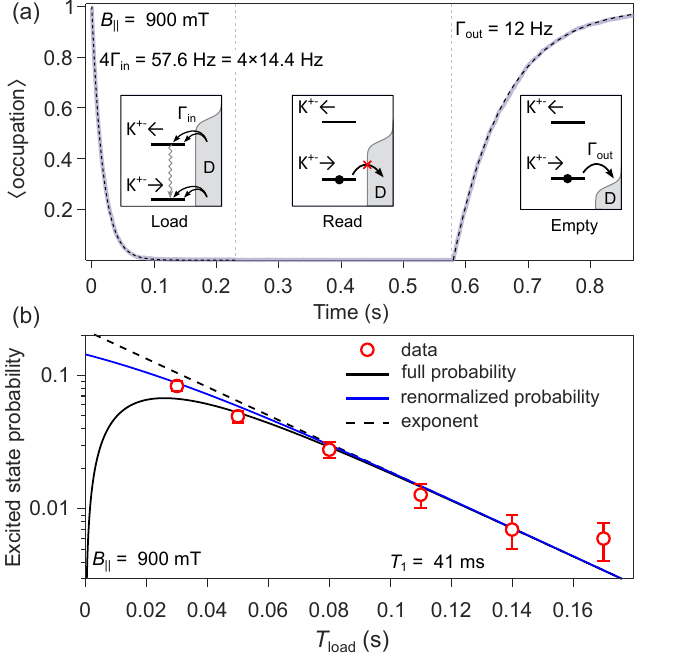}
	\caption{\textbf{Calculation of excited state probability.}~(a)~Averaged over approximately 2000 shots, the occupation of the QD is shown as a function of time for $B_{||}=\SI{900}{\m T}$ and a loading time $T_{\mathrm{load}}=\SI{220}{\m s}$, significantly longer than $T_{1}=\SI{41}{\m s}$. Dashed lines represent exponential fits with outlined tunneling rates $\Gamma_{\mathrm{in/out}}$. Energy diagrams sketch the standard three-level spin readout protocol. (b)~The excited state probability is plotted against the loading time $T_{\mathrm{load}}$. The dashed line shows an exponential fit with $T_{1}=\SI{41}{\m s}$. The black line represents the full analytical probability, considering extracted tunneling rates and $T_{1}=\SI{41}{\m s}$. The blue solid line depicts the renormalized probability after excluding all non-loaded shots.}
	\label{sup_fig_2}
\end{figure}

\clearpage
\newpage

\begin{figure*}[!h]
	\includegraphics[width=0.5\columnwidth]{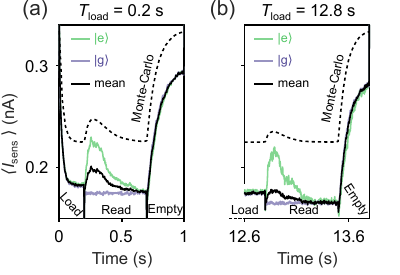}
	\caption{\textbf{Dark counts.}~Average of single-shot sensor current traces identified as the ES (green), the GS (purple), and the average of both (black) are shown for two very distinct loading times $T_{\mathrm{load}}=0.2~\mathrm{s}$~(a) and $12.8~\mathrm{s}$~(b) measured in $B_{\perp}=55~\mathrm{mT}$. Dashed lines show the mean of the traces generated via Monte Carlo simulations. The vertical offset for the simulated data is introduced for clarity. }
	\label{fig_6}
\end{figure*}

\clearpage
\newpage

\begin{figure*}[h!]
	\includegraphics[width=0.7\columnwidth]{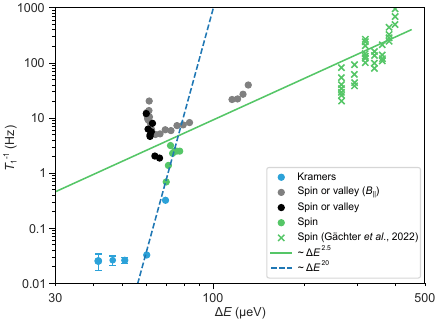}
	\caption{\textbf{Ruling out pure spin relaxation.}~Measured inverse relaxation time $T_{1}^{-1}$ as a function of the energy splitting between relevant states. Circular markers of different colors correspond to different relaxation channels as outlined in the legend. Crosses correspond to the data points from~\cite{PRXQuantum.3.020343}, measured with the same Elzerman technique in a weakly coupled single BLG QD. Solid line shows the $\sim\Delta E^{2.5}$ trend in spin relaxation data. Dashed line shows $\sim\Delta E^{20}$, ruling out the assumption that Kramers qubit relaxation is dominated by the spin channel. }
	\label{fig_5}
\end{figure*}

\clearpage
\onecolumngrid

\begin{center}
	\Large{\textbf{Supplementary Information}}
\end{center}

	% Set the distance between line numbers and text
	\setlength\linenumbersep{0.2em} 
	
	% Enable line numbers and place them on the outer edge of each column
	\modulolinenumbers[1] % Number lines in steps of 5 (optional)
	%\linenumbers
	%\switchlinenumbers

	\maketitle
	
%	\tableofcontents
	
	%\newpage
	\section{Tunneling rate spectroscopy}
	
	\begin{figure*}[tbh!]
		\includegraphics[width=1\columnwidth]{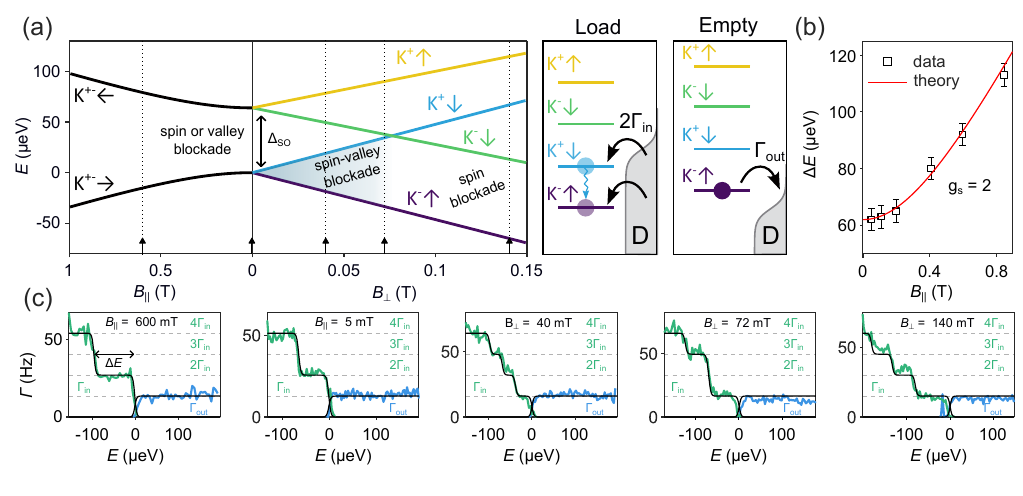}
		\caption{\textbf{Tunneling rate spectroscopy.}~(a)~Energy spectrum of a single carrier in the BLG QD plotted as a function of in-plane and out-of-plane magnetic fields. Two-level pulse sequence for determining the in- and out-tunneling rates. The tunneling-in rate is proportional to the number of available energy states, while the tunneling-out rate is independent of the carrier's state. (b)~Energy splitting between Kramers pairs $\Delta E$ is plotted as a function of in-plane magnetic field $B_{||}$. Solid red line is a theoretical curve with $g_{\mathrm{s}}=2$ and $\Delta_{\mathrm{SO}}=62~\mathrm{\mu eV}$. Data points are presented as mean values $\pm$ the thermal broadening $3.5 k_{\mathrm{B}}T_{e}/2 \approx~4~\mathrm{\mu eV}$ of the Fermi-Dirac function. (c)~Measured at different magnetic fields, tunneling-in (green) and -out (blue) rates are plotted as a function of the energy shift $E$ relative to the GS. Black solid lines show theoretical curves with the sum of Fermi-Dirac distribution functions.}
		\label{sup_fig_0}
	\end{figure*}

	Tunneling rate spectroscopy technique assumes that all four states share the same orbital wave function. Consequently, the change in tunneling-in rates can be viewed as consecutive step functions, where the number of steps corresponds to the number of available states. To effectively measure the tunneling rates, we apply slow ($1-3~\mathrm{Hz}$) periodic two-level square pulses which aim to `Load' and `Empty' the dot in each cycle as sketched in Fig.~\ref{sup_fig_0}a. By measuring the waiting time $t$ between the pulse and the actual tunneling event, we can extract both tunneling rates by fitting the exponential distribution $\sim \exp{(-\Gamma_{\mathrm{in(out)}}t)}$. Additional details can be found in~\cite{Duprez2024}, where similar measurements were conducted on the same device. Fig.~\ref{sup_fig_0}a shows the spectrum of the single hole in BLG QD as a function of in- and out-of-plane magnetic fields~\cite{Knothe_2022}. Close to zero magnetic field, two doubly-degenerate Kramers pairs manifest themselves as two plateaus in measured tunneling-in rate at approximately $\Gamma_{\mathrm{in}}=2\Gamma_{\mathrm{out}}$ and $\Gamma_{\mathrm{in}}=4\Gamma_{\mathrm{out}}$ as shown in Fig.~\ref{sup_fig_0}c for $B_{||}=\SI{5}{\m T}$. As expected, the tunneling-out rate is constant $\Gamma_{\mathrm{out}}\approx\SI{13}{Hz}$ far away from the transition. The width of the $2\Gamma_{\mathrm{in}}$ step at zero magnetic field corresponds to the spin--orbit gap $\Delta_{\mathrm{SO}}$. As we increase the in-plane field to $B_{||}=\SI{600}{\m T}$, the degeneracy of the doublets stays the same, while the energy gap is increasing according to the in-plane polarization of the spin $\Delta=\sqrt{\Delta_{\mathrm{SO}}^{2}+(g_{\mathrm{s}}\mu B_{||})^{2}}$. We fit the data with a sum of Fermi-Dirac distribution functions~\cite{Duprez2024} with electron temperature $T_{\mathrm{e}}=\SI{25}{\m K}$ and $\Delta_{\mathrm{SO}}=\SI{62}{\micro eV}$ and $g_{\mathrm{v}}=14.5$. In contrast, the perpendicular magnetic field disrupts the degeneracy of the doublets. At $B_{\perp}=\SI{40}{\m T}$, we notice four nearly evenly spaced energy steps, each with $\Gamma_{\mathrm{in}}=\Gamma_{\mathrm{out}}$. As we increase the field, at $B_{\perp}=\SI{80}{\m T}$, two ESs intersect, leading to the merging of the second and third plateaus into a single one with $\Gamma_{\mathrm{in}}=3\Gamma_{\mathrm{out}}$ as evident from the data. However, beyond the crossing point, at $B_{\perp}=\SI{140}{\m T}$, the second plateau reappears with $\Gamma_{\mathrm{in}}=2\Gamma_{\mathrm{out}}$. Note that we were only able to tune the dot to this textbook spectrum when the tunneling rates were as low as $10-20~\mathrm{Hz}$.

	\section{Spectroscopy around relaxation hotspots}
	
	\begin{figure*}[tbh!]
		\includegraphics[width=1\columnwidth]{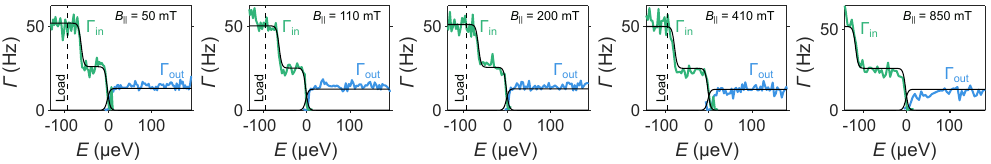}
		\caption{\textbf{Tunneling rate spectroscopy around relaxation hotspot.}~Measured at different in-plane magnetic fields $B_{||}$, tunneling-in (green) and -out (blue) rates are plotted as a function of the energy shift $E$ relative to the GS. Black solid lines show analytical curves with the sum of Fermi-Dirac distribution functions, $g_{s}=2$ and $\Delta_{\mathrm{SO}}=60~\mathrm{\mu eV}$. The loading levels are marked with the dashed lines.}
		\label{sup_fig_3}
	\end{figure*}
	We noticed a sharp but rather shallow (2-3 times drop) hotspot in the measured $T_1$ around $B_{||}\approx\SI{110}{\m T}$. In Fig.~\ref{sup_fig_3}, we show the results of the tunneling rate spectroscopy around this point. As expected, all tunneling-in rate curves exhibit a similar two-step behavior as a function of the energy shift $E$ in respect to the Fermi level in the leads. The width of the first step is defined by the splitting between two Kramers pairs and matches well with the theoretical predictions $\Delta E~=~\sqrt{\Delta_{\mathrm{SO}}^{2}+(g_{\mathrm{s}}\mu_{\mathrm{B}}B_{||})^2}$, where $\Delta_{\mathrm{SO}}=\SI{60}{\micro eV}$ and $g_{\mathrm{s}}=2$. The only thing we can point out at $B_{||}=\SI{110}{\m T}$ (right on the hotspot) is a sign of a higher excited state emerging as an additional step in $\Gamma_{\mathrm{in}}$ at $E<-\SI{100}{\micro eV}$. However, this state is lying higher than the loading level, outlined by the dashed line, and should not contribute to the measurements.

	\textcolor{mycolor}{\section{Coherent manipulation of the Kramers states}}
	
	\textcolor{mycolor}{\subsection{Kramers singlet-triplet qubit}}
	
	\textcolor{mycolor}{For a double quantum dot with a single carrier in each dot (1,1), the ground state is any combination of the following 4 states: $|K^{+}\downarrow,K^{+}\downarrow\rangle$,~$|K^{-}\uparrow,K^{-}\uparrow\rangle$,~$|K^{+}\downarrow,K^{-}\uparrow\rangle$,~$|K^{-}\uparrow,K^{+}\downarrow\rangle$.}
	
	\textcolor{mycolor}{In the singlet-triplet basis:
		\begin{align*}
			&|A\rangle(1,1)=\frac{1}{\sqrt{2}}\big(\textcolor{blue}{|K^{+}\downarrow,K^{-}\uparrow\rangle-|K^{-}\uparrow,K^{+}\downarrow\rangle}\big)\\
			&|B_{0}\rangle(1,1)=\frac{1}{\sqrt{2}}\big(|K^{+}\downarrow,K^{-}\uparrow\rangle+|K^{-}\uparrow,K^{+}\downarrow\rangle\big)\\
			&|B_{+}\rangle(1,1)=|K^{+}\downarrow,K^{+}\downarrow\rangle\\
			&|B_{-}\rangle(1,1)=|K^{-}\uparrow,K^{-}\uparrow\rangle
	\end{align*}}

	%\begin{figure*}[t!]
	%	\includegraphics[width=0.85\columnwidth]{sup_fig4_v1.pdf}
	%\caption{\textcolor{mycolor}{Kramers singlet-triplet qubit. (a)~Tunnel coupling between %%(1,1) and (2,0) orbital states. (b)~Energies of the two-electron spin-valley singlet and %triplet levels in a double dot as a function of detuning between the levels in the two dots at $B=0$.}}
	%\label{sup_fig_4}
	%\end{figure*}

	\textcolor{mycolor}{In the case of two-particles in a single dot, the ground state at B=0  has been shown to be a triple-degenerate valley-singlet, spin-triplet state~\cite{PhysRevLett.123.026803}:}
	
	\textcolor{mycolor}{\begin{align*}
			&|S^{\mathrm{v}}T^{\mathrm{s}}_{0}\rangle(2,0)=\frac{1}{{2}}\big(|K^{+}\uparrow,K^{-}\downarrow\rangle-|K^{-}\downarrow,K^{+}\uparrow\rangle+\textcolor{blue}{|K^{+}\downarrow,K^{-}\uparrow\rangle-|K^{-}\uparrow,K^{+}\downarrow\rangle}\big)\\
			&|S^{\mathrm{v}}T^{\mathrm{s}}_{+}\rangle(2,0)=\frac{1}{\sqrt{2}}\big(|K^{+}\uparrow,K^{-}\uparrow\rangle-|K^{-}\uparrow,K^{+}\uparrow\rangle\big)\\
			&|S^{\mathrm{v}}T^{\mathrm{s}}_{-}\rangle(2,0)=\frac{1}{\sqrt{2}}\big(|K^{+}\downarrow,K^{-}\downarrow\rangle-|K^{-}\downarrow,K^{+}\downarrow\rangle\big)\\
	\end{align*}}
	
	\textcolor{mycolor}{As we see, only $|S^{\mathrm{v}}T^{\mathrm{s}}_{0}\rangle(2,0)$ and $|A\rangle(1,1)$ (highlighted in blue) can be coupled by interdot tunnelling in the assumption that the spin and valley quantum numbers are conserved during the tunnelling process. 
		%The resulting energy level diagram is shown in Fig.~\ref{sup_fig_4}b as a function of detuning %$\delta$ between (1,1) and (2,0) charge states. 
		One can show that the resulting tunneling Hamiltonian, spanned by the Kramers singlet $|A\rangle$(1,1) and unpolarized triplet $|B_{0}\rangle(1,1)$ states is similar to its spin analog~\cite{doi:10.1126/science.1116955}. 
		%$$H=\frac{1}{2}J(\delta)\sigma_{\mathrm{z}}+\frac{1}{2}\Delta %g_{\mathrm{v}}\mu_{\mathrm{B}}B_{\perp}\sigma_{\mathrm{x}} $$
		%Here the second term arises from the difference in valley g-factors between the two dots. In the %following section, we demonstrate that $g_{\mathrm{v}}$ can be highly tunable by via gate %voltages. Magnetic (ESR-like) driving of the qubit can be achieved by modulating $B_{\perp}(t)$ %with a microwave antenna or micromagnet. Alternatively, electrical driving can be realized by %modulating $\Delta g_{\mathrm{v}}(t)$ using barrier gate voltages and $J(\delta)$ using middle %barrier gate. Additionally, standard non-adiabatic pulsing across the detuning axis can be %implemented in a manner similar to spin singlet-triplet qubits in %GaAs~\cite{doi:10.1126/science.1116955} or Ge \cite{Jirovec2021}. The readout of the Kramers %singlet-triplet qubit can be achieved via Pauli spin-valley %blockade~\cite{PhysRevLett.128.067702}, which is very similar to readout methods used in spin %systems such as GaAs.
	}

	\textcolor{mycolor}{\subsection{Loss-DiVincenzo Kramers qubit via finite intervalley mixing}}
	
	\textcolor{mycolor}{In the presence of a finite valley scattering term, the operation of the Kramers qubit is relatively straightforward, as theoretically described in~\cite{PhysRevB.81.195418} as well as experimentally demonstrated on carbon nanotubes by Laird et al.~\cite{Laird2013}. Unlike CNTs, however, $\Delta_{K^{+}K^{-}}<0.5~\mathrm{MHz}$ in BLG is significantly suppressed due to the low long- and short-range disorder which would otherwise cause large-momentum scattering. Following the theoretical work on CNTs, in a presence of the finite intervalley mixing, we can project our original single particle Hamiltonian:
		$$H=-\frac{1}{2}\big(\tau_{\mathrm{z}}\Delta_{\mathrm{SO}}\sigma_{\mathrm{z}}+\tau_{\mathrm{x}}\Delta_{K^{+}K^{-}}   \big)+g_{\mathrm{s}}\mu_{\mathrm{B}}\Vec{\sigma}\cdot\Vec{B}+\tau_{\mathrm{z}}g_{\mathrm{v}}\mu_{\mathrm{B}}B_{\perp} $$
		onto the lowest two eigenstates (Kramers pair) yielding an effective spin-1/2 system:}
	
	\textcolor{mycolor}{$$H^{*}=\frac{1}{2}\mu_{B}\big( s_{\mathrm{z}}g_{\perp}B_{\perp}  + s_{\mathrm{x}}g_{||}B_{||} \big),~~\mathrm{where}~g_{\perp}=g_{\mathrm{s}}+2g_{\mathrm{v}}\frac{\Delta_{\mathrm{SO}}}{\sqrt{\Delta_{\mathrm{SO}}^{2}+\Delta_{K^{+}K^{-}}^{2}}},~~g_{||}=2g_{\mathrm{s}}\frac{\Delta_{K^{+}K^{-}}}{\sqrt{\Delta_{\mathrm{SO}}^{2}+\Delta_{K^{+}K^{-}}^{2}}}$$}
	
\begin{figure*}[t!]
		\includegraphics[width=1\columnwidth]{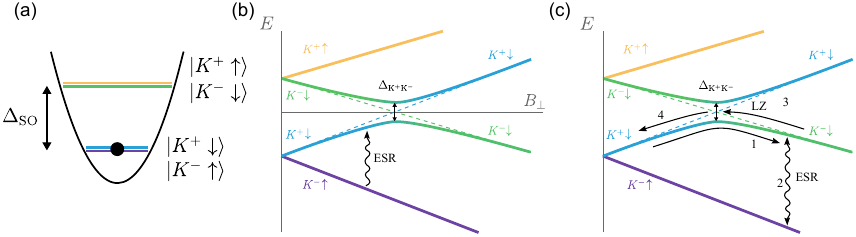}
		\caption{\textcolor{mycolor}{\textbf{Kramers qubit manipulation based on valley-mixing.}~(a)~Single carrier QD in BLG. (b) Energy dispersion of a single carrier in BLG QD in the presence of a finite intervalley scattering term $\Delta_{K^{+}K^{-}}$. The Kramers qubit can be driven via Electric Spin Resonance (ESR).}}
		\label{sup_fig_5}
	\end{figure*}

	\textcolor{mycolor}{Here, $s_\mathrm{z}$ and $s_\mathrm{z}$ are spin Pauli matrices, and $B_{||}$ represents the in-plane magnetic field parallel to the BLG plane. The above Hamiltonian allows for the use of conventional ESR to drive (see Fig.~\ref{sup_fig_5}b) the qubits, either by employing a microwave antenna or micromagnet, or by utilizing the geometric bends of the QD, as demonstrated in CNTs~\cite{Laird2013}. Note that the desirable in-plane g-factor component $g_{||}$, is expected to vanish as $\Delta_{\mathrm{K^{+}}\mathrm{K^{-}}}\to0$. This originates from the mixing of the first $K^{+}\downarrow$  and second $K^{-}\downarrow$ excited states as we approach the anticrossing point, as shown in Fig.~\ref{sup_fig_5}b. In BLG,  $\Delta_{\mathrm{K^{+}}\mathrm{K^{-}}}\ll \Delta_{\mathrm{SO}}$, so $g_{||}\ll g_{\mathrm{s}}$ and as a result, Rabi frequencies are expected to be significantly slower than those of spin Loss-DiVincenzo qubits~\cite{RevModPhys.95.025003}.}
	
	\textcolor{mycolor}{\subsection{Valley qubit via finite intervalley mixing}}
	
	\textcolor{mycolor}{Another interesting possibility is to drive a pure valley qubit using a non-adiabatic pulsing across the Landau-Zener transition shown in Fig.~\ref{sup_fig_5}c. Fast pulsing across the magnetic field axis can be achieved by leveraging the tunable valley g-factor. One possible protocols is illustrated in Fig.~\ref{sup_fig_5}c:\\
		\begin{enumerate}
			\item Adiabatically drive the system across the LZ transition, to obtain a quantum state encoded solely in the spin, and perform Elzerman readout to initialize the spin in the ground state.
			\item Apply a spin $\pi$-pulse using ESR drive to excite the carrier into $K^{-}\downarrow$ state.
			\item Non-adiabatically pulse close to the anticrossing and wait for Larmor precession of the valley with rate equal to $\Delta_{\mathrm{K}^{+}\mathrm{K}^{-}}$.
			\item Non-adiabatically pulse far from the anticrossing and perform energy-selective readout between $K^{-}\downarrow$ and $K^{+}\downarrow$ states.
	\end{enumerate}}
	%\newpage
	
\begin{figure*}[tbh!]
		\includegraphics[width=1\columnwidth]{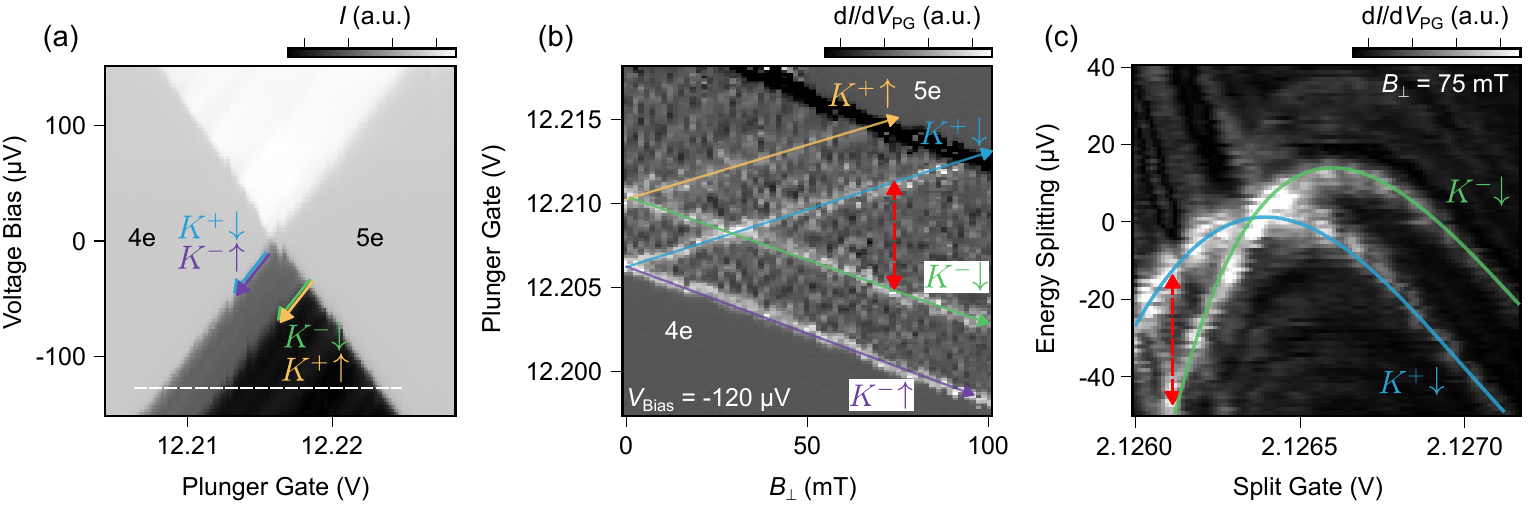}
		\caption{\textcolor{mycolor}{\textbf{Tunable valley magnetic moment.}~(a)~The Coulomb diamond measured in the direct current through the quantum dot (QD) between the 4-to-5 electrons transition. The white dashed line indicates the bias voltage used for spectroscopy in a magnetic field. (b)~Finite-bias spectroscopy as a function of the out-of-plane magnetic field shows the canonical spectrum. (c)~The energy splitting between the first and second excited states is presented as a function of the split gate, measured at a fixed magnetic field of $B_{\perp}=75~\mathrm{mT}$. The reverse crossing of the states indicates the high tunability of the valley g-factor. Solid lines are provided as guides for the eye.}}
		\label{sup_fig_6}
	\end{figure*}

	\textcolor{mycolor}{\section{Gate-tunable valley magnetic moment}}
	\textcolor{mycolor}{The unique property of BLG is a gate-tunable valley g-factor, which allows electrical driving of the Kramers singlet-triplet states as well as adiabatic sweeps across the intervalley (anti)crossing. Here we experimentally demonstrate the tunability of $g_\mathrm{v}$ in the range from approximately 10 to 40 by simply varying the split-gate voltage. In Fig.~\ref{sup_fig_6} we measured the transition between 4 and 5 electrons in the QD, which behave similarly to the 0-to-1 transition, as 4 carriers form a full orbital shell. We did not observe this effect for the 0-to-1 transition within the range of split-gate voltages where the excited states of the dot are still resolvable. We attribute this to the fact that a quantum dot with 4 electrons is inherently larger than one starting with 0 electrons, and the size is what defines the valley magnetic moment~\cite{doi:10.1021/acs.nanolett.0c04343}.}
	
	\textcolor{mycolor}{In Fig.~\ref{sup_fig_6}a, a Coulomb diamond is measured via direct current through the QD (high tunnelling rates). Finite-bias spectroscopy in Fig.~\ref{sup_fig_6}b reveals the canonical single-particle energy spectrum as a function of the out-of-plane magnetic field, including two Kramers pairs. At fixed perpendicular magnetic field $B_{\perp}~=~75~\mathrm{mT}$, we track the energy splitting (red arrow in Fig.~\ref{sup_fig_6}b) between $K^{-}\downarrow$ (green) and $K^{+}\downarrow$ (blue) states as a function of split-gate voltage as shown in Fig.~\ref{sup_fig_6}c. To directly measure the energy splitting, we sweep the bias voltage rather than the plunger gate, avoiding the need to separately measure the lever arm for each split-gate voltage. We can clearly observe the two excited states moving closer together and eventually crossing as the split-gate voltage is slightly increased ($1.5~\mathrm{mV}$). This is directly related to the change of valley g-factor and moving of the crossing point between $K^{-}\downarrow$ (green) and $K^{+}\downarrow$ (blue) states to higher magnetic fields in Fig.~\ref{sup_fig_6}b (from $~30~\mathrm{mT}$ to $115~\mathrm{mT}$). The position of this crossing is defined only by the spin-orbit gap $\Delta_{\mathrm{SO}}$ and valley g-factor $g_{\mathrm{v}}$, while we ensure that the SO gap remains unchanged as a function of the split-gate voltage. This tunability enables nonadiabatic pulsing across the crossing point using electrical signals as well as ESR-like driving of the singlet-triplet qubit. Since the split-gate also influences the transition in plunger gate voltage, we virtually compensate for the plunger gate voltage to stay always around the the transition in Fig.~\ref{sup_fig_6}c. This linear compensation is responsible for non-monotonic trend of excited state lines.} 
	
	\newpage
	\section{Monte Carlo simulations }

	\subsection{Single-shot readout in a three-state quantum dot}
	In our simulation, we considered the quantum dot to be a 3-level system, with ground state, first excited state, and second excited state. The whole process of readout is divided into 4 parts - the loading stage, the waiting stage, the measurement stage, and the unloading stage.
	
	In the loading stage, a charge carrier is loaded into the quantum dot. Because all the 3 states are well below the electrochemical potential of the leads, we assume that the charge carrier has an equal chance of ending up in either one of the 3 states.
	
	If the charge carrier ends up in the ground state, it will remain there during the waiting stage. If the charge carrier is in one of the excited states, it might relax to a lower-energy state.
	
	In the measurement stage, the energy level of the states is moved up by applying a voltage pulse. If the charge carrier is in the ground state, it may tunnel out. If it is in the first excited state, it may either tunnel out or relax to the ground state, and if it is in the second excited state, it may tunnel out, relax to the ground state, or relax to the first excited state. If the dot is empty at any point during the measurement stage, another charge carrier from the leads may fill either one of the 3 states, however, this time with different probabilities.
	
	In the unloading stage, the energy level of the states is moved up way above the electrochemical potential of the leads, which causes the charge carrier to tunnel out from the quantum dot.
	
	\subsection{Probability of tunneling}
	
	All tunneling processes mentioned in the section above are probabilistic and described through tunneling rates. In the loading stage, we use tunneling rate $3 \Gamma_{\mathrm{in},0}$, because the electron can tunnel into either one of the 3 states. In the unloading stage, we use tunneling rate $\Gamma_{\mathrm{out},0}$. In the simulation, we used the value (taken from the experiments) $\Gamma_{\mathrm{in},0}$=$\Gamma_{\mathrm{out},0}=\SI{15}{Hz}$. 
	
	In the measurement stage, the tunneling rate can be calculated using Fermi-Dirac distribution:
	
	\begin{equation}
		\Gamma_{\text{in/out}}=\frac{\Gamma_{\text{in/out},0}}{1+e^{-\frac{E}{k_{\mathrm{B}} T}}},
	\end{equation}
	
	where $T$ is temperature (in the simulation we used value $T=\SI{25}{\m K}$), $k_B$ is the Boltzmann constant and $E$ is the energy of the given state, assuming that the electrochemical potential of the leads corresponds to $E=0$. We can notice that if the energy level of the given state is a bit below the electrochemical potential of the leads, the tunneling rate $\Gamma_{\text{in}}$ is much larger than $\Gamma_{\text{out}}$ and therefore it is more probable for a charge carrier to tunnel in than the opposite situation.

	To calculate the probability of a charge carrier tunneling either in or out of the dot, we need to use Bayes' theorem. It states that the conditional probability of an event $A$ happening while event $B$ is true can be calculated using the following equation: 
	
	\begin{equation}
		P(A|B) = \frac{P(B|A) \cdot P(A)}{P(B)}.
		\label{Bayes}
	\end{equation}
	
	We can apply this theorem to calculate the probability of the dot getting occupied in time interval $[t, t+\Delta t]$, provided that the dot has been empty before that, in time interval $[0, t]$:

	\begin{equation}
		\begin{split}
			&P(\mathrm{dot\: got\: occupied\: in\: [t, t+\Delta t]}| \mathrm{dot \: was \: empty \: in\: [0, t]})=  \\
			=& \frac{P(\mathrm{dot\: got\: occupied\: in\: [t, t+\Delta t]}) \cdot P(\mathrm{dot \: was \: empty \: in\: [0, t]}|\mathrm{dot\: got\: occupied\: in\: [t, t+\Delta t]})}{P(\mathrm{dot \: was \: empty \: in\: [0, t]})} \\[15pt]
			=& \frac{\int_t^{t+\Delta t} \Gamma e^{-\Gamma t} \, dt}{1-\int_0^t \Gamma e^{-\Gamma t} \, dt} = \frac{e^{-\Gamma t}-e^{-\Gamma(t+\Delta t)}}{e^{-\Gamma t}} = 1 - e^{-\Gamma \Delta t}.
		\end{split}
		\label{threshold}
		\end{equation}
%	\onecolumngrid
	In the calculation, we used the fact that the probability density of a tunneling event at a time $t$ is given by
	
	\begin{equation}
		p(t)\,dt = \Gamma e^{-\Gamma t} dt.
	\end{equation}
	
	In our simulation, we discretized the whole readout process, with a time step $t_{\text{step}}=0.0005\, \text{s}$. In each step, we generate a random number from 0 to 1 and calculate the probability given by equation \eqref{threshold}. If the randomly generated number is smaller than the calculated probability, the given tunneling event occurs.

%	\newpage
	\textcolor{mycolor}{\section{Charge stability diagram}}
	
	\begin{figure*}[tbh!]
		\includegraphics[width=0.75\columnwidth]{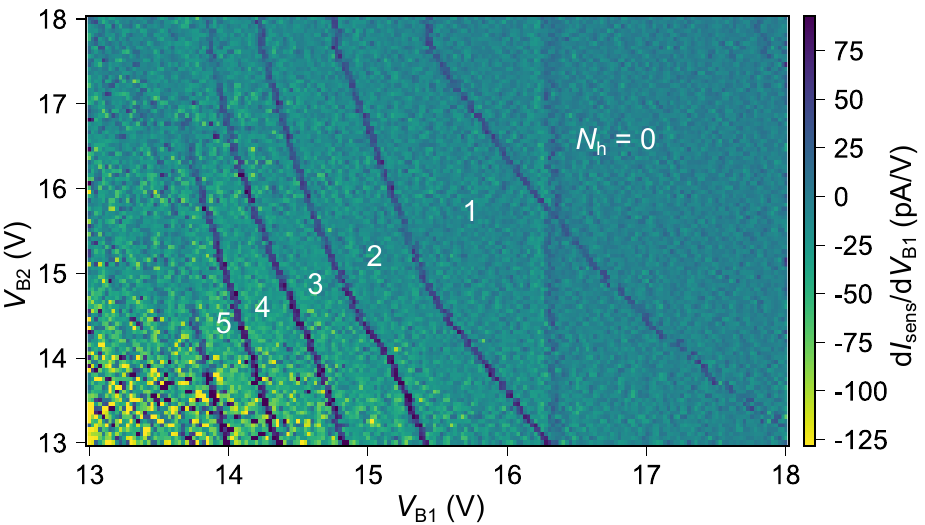}
		\caption{\textcolor{mycolor}{\textbf{Barrier-barrier charge stability diagram.}~Additional charge stability diagram. Sensor current derivative as a function of the barrier gate voltages. $N_{\mathrm{h}}$ indicates the number of hole carriers inside the QD.}}
		\label{sup_fig_7}
	\end{figure*}

%	\bibliography{bibliography_Kramers}

\end{document}